# Acoustic Higher-Order Weyl Semimetal with Bound Hinge States in the Continuum

Zhenhang Pu[1], Hailong He[1*], Licheng Luo[1], Qiyun Ma[1], Liping Ye[1], Manzhu Ke[1*], and Zhengyou Liu[1,2*]

1 Key Laboratory of Artificial Micro- and Nano-structures of Ministry of Education and School of Physics and Technology, Wuhan University, Wuhan 430072, China.

2 Institute for Advanced Studies, Wuhan University, Wuhan 430072, China.

*Corresponding authors: hehailong@whu.edu.cn; mzke@whu.edu.cn; zyliu@whu.edu.cn

**Abstract**

Higher-order topological phases have raised widespread interest in recent years with the occurrence of the topological boundary states of dimension two or more less than that of the system bulk. The higher-order topological states have been verified in gapped phases, in a wide variety of systems, such as photonic and acoustic systems, and recently also observed in gapless semimetal phase, such as Weyl and Dirac phases, in systems alike. The higher-order topology is signaled by the hinge states emerging in the common bandgaps of the bulk states and the surface states. In this Letter, we report our first prediction and observation of a new type of hinge states, the bound hinge states in the continuum (BHICs) bulk band, in a higher-order Weyl semimetal implemented in phononic crystal. In contrast to the hinge state in gap, which is characterized by the bulk polarization, the BHIC is identified by the nontrivial surface polarization. The finding of the topological BHICs broadens our insight to the topological states, and may stimulate similar researches in other systems such as electronic, photonic, and cold atoms systems. Our work may pave the way toward high-Q acoustic devices in application.

*Introduction.*—There has been a growing interest in higher-order topology over recent years. Higher-order topology means that a system hosts topological boundary states with dimension two or more less than that of the system itself, or explicitly, a two-dimensional (2D) system of 0D corner states, or a 3D system of 0D corner or 1D hinge states [1-4]. Such bulk-boundary correspondence extends the one in the conventional or first-order topology where the boundary states have dimension one less than that of the bulk. Higher-order topology was first predicted and manifested in topological insulators [1-13] hosting topological corner states or hinge states, and most recently also in 3D topological semimetals including the Dirac [14,15], Weyl [16,17] and nodal line semimetals [18,19]. Unlike the conventional topological semimetals identified by the Fermi arc surface states, the higher-order semimetals are characterized by hinge states on their 1D boundaries. Benefiting from the macroscopic scale and the flexibility in design of classical artificial structures, higher-order semimetals, since they were proposed [4,14], were first experimentally implemented in acoustic systems [20-23], and then in photonic crystals [24], electric circuits [18] and so on. In particular, for higher-order Weyl semimetals (HOWSMs) realized in phononic crystals (PCs), it has been demonstrated that the hinge states emerge in the common gaps of the bulk and surface states connecting the projected Weyl points (WPs) on the hinge Brillouin zones (BZs) [20,21].

As localized states (in full dimensions or partial dimensions), corner states or hinge states generally exist in the gaps of the bulk bands. But spectral isolation is not a must for the emergence of localized states. It has been shown that the corner states can also appear in the bulk bands [11,25-27], revealing as the bound states in the continuum (BICs) [28-31]. Embedded in the continuous spectrum of the radiating bulk states but remaining intact or without coupling with them [30,31], BICs may enable the high Q factors in relevant applications [32-34]. Since only the BICs of the corner states [11,25-27] are observed in higher-order topological materials, a natural question can be raised: can the hinge states, extended along the hinge but localized transversely, appear to be the BHICs in topological insulators or semimetals?

In this Letter, we theoretically predict and experimentally verify the existence of BHICs in a HOWSM. Our 3D model, developed from the $\alpha$-$SiO_2$ structure [35], hosts two sets of triangular Weyl complexes (TWCs) in its three contiguous bands; each set consists of a quadratic WP and two ordinary WPs, and has a net charge of zero. It is found that each TWC sustains a branch of the conventional hinge states in gap (HIGs), but more interestingly, in-between the two TWCs and in the middle bulk band two novel BHIC bands appear, extending over almost the whole hinge BZ. Different from the usual HIGs, which are identified by the bulk polarization, the BHICs are the consequence of the non-zero surface polarization. The three WPs within each TWC can be tuned to identical energy as the ideal WPs [36,37], benefiting the study for Weyl physics. With the model implemented in a PC, all results are corroborated by both the simulations and experiments. Our work not only provides further insight into the topological states but also paves the way toward state of the art high-Q devices in acoustics.

*Tight-binding model.*—Our lattice model is developed based on the structure of the $\alpha\text{-}SiO_2$, which is recently reported to host the TWCs for phonons [35]. To simplify the structure, we remove all the oxygen atoms and allow the silicon atoms to couple to the next-nearest neighbors. As such, we derive our lattice model as shown in Fig. 1(a), viewed from the $z$ direction, with the colored spheres (red, turquoise and blue) indicating the atoms at different $z$ positions, related by three-fold screw axes locatedat the centers of the triangles and the hexagons, respectively. Obviously, the bulk crystal is a Kagome lattice but with the lattice sites related by screw symmetry along $z$ direction. $a_0$ and $h_0$ respectively being the in-plane and out-of-plane lattice constants, each unit cell (green hexagon) contains three identical sites, i.e. the differently colored spheres, as given in Fig. 1(b) with an amplified view. With the staggered nearest-neighbor hopping strengths denoted by $t_0$ (orange) and $t_1$ (silver), and the next-nearest-neighbor hopping by $t_{nnn}$ (black), three energy bands can be read off from the Bloch Hamiltonian (see Ref. [38]), as shown in Fig. 1(c). It is noteworthy that, the degenerate nodes between bands 1 and 2 (or bands 2 and 3) possess one quadratic WP carrying a topological charge $+2$ (or $-2$) pinned to the time-reversal-invariant momenta $\Gamma$ (or A), and two ordinary WPs with topological charge $-1$ (or $+1$) located at $K'_1$ and $K_2$ (or $K_1$ and $K'_2$), thus reproducing the double TWCs in $\alpha\text{-}SiO_2$ except for the differences in the positions of the charge-1 ($+1$ or $-1$, except specified explicitly hereafter) WPs. However, in our reduced model, the three WPs in each TWC have been tuned to identical energy, or in other words, they are ideal WPs. The ideal WPs can provide a clean energy window without containing any other nodes, for studying pure Weyl physics. The two sets of the TWCs in the first BZ in green ($E_1 = -1.10$) and magenta ($E_2 = 1.10$), with three WPs each, are displayed in the left panels of Figs. 1(d) and 1(e), respectively. By adjusting the hopping strengths, the positions of the charge-1 WPs in the $k_z$ direction in reciprocal space, the dispersive features of the quadratic ones [37,39,40], and the energy separation of the WPs within each TWC, can all be engineered (see Ref. [38]).

To characterize the bulk topological properties of our model, we investigate the first- (Chern number $C$) and second-order (bulk polarization $p_b$) bulk topological indexes by viewing our 3D system as a set of $k_z$-dependent 2D subsystems [16,20-23]. These subsystems possess two bandgaps except those containing WPs. The $k_z$-dependent topological indexes for the two gaps are given respectively by Figs. 1(d) and 1(e), with the middle (right) panels denoting the Chern numbers $C$ (polarizations $p_b$) [1,2]. For the lower bandgap [see Fig. 1(d)], a non-zero Chern number with $C = \pm 1$ emerges in the region $0 < |k_z| < 0.5\pi/h_0$ (which exactly corresponds to the projection of the first TWC on $k_z$), while a nontrivial $p_b = (1/2, 1/2\sqrt{3})$ (and a vanishing Chern number $C$) appear in the region of $|k_z| > 0.5\,\pi/h_0$ but within the projected BZ in the $z$ direction. But for the upper bandgap, the indexes in the corresponding regions are reversed [see Fig. 1(e)]. Specifically, if we construct a parallelogram-shaped prism structure as shown in Fig. 1(a), the nontrivial bulk polarizations $p_b$ gives rise to hinge states localized at the lower left acute hinge [4,11,12], which identify the HOWSM. Throughout this Letter, we will only focus on this hinge boundary and its two adjacent surfaces.

To verify the existence of surface states and hinge states, we calculate surface- and hinge-projected dispersions. In Fig. 1(f), we demonstrate the surface dispersions of the XZ surface of a ribbon structure [dashed rectangle in Fig. 1(a)], and only the results for $k_z \geq 0$ are displayed, since those for $k_z \leq 0$ are the time-reversal of them. It is worth noting that the Fermi arcs (green and magenta lines), starting from the quadratic WPs and ending at the charge-1 ones, are not shielded by any bulk states, exhibiting particular advantage of the ideal Weyl semimetal in surface transport. When projected to the $k_z$ direction, as shown in Fig. 1(g), the surface states (in yellow) partly overspread the gaps of the bulk states, and partly overlap with the bulk states (in grey) [20,41]. Hinge states along the $z$ direction of the prism structure in Fig. 1(a) are also displayed in the figure. As expected, we observe the HIGs (red lines) connecting the charge-1 WPs in bulk gaps, as indicated by the corresponding nontrivial bulk polarization $p_b$. Interestingly, two extra branches of hinge states, or BHICs (cyan line) appear, which are embedded in the middle bulk band but situated in the surface bandgaps, extending over almost the entire hinge BZ.

In order to figure out the origin of the unique BHICs, we explore the topological properties of the surface state bands. Actually, the surface states can be considered as a new two-band subspace with gap closing at $(k_x, k_z) = (0, \pm 0.5\,\pi/h_0)$ [yellow spheres in Fig. 1(f)]. In Fig. 1(h), we show the surface polarization $p_s$ along the $x$ direction in the surface bandgaps, formulated in terms of the Berry phase [14,42,43], and find it non-zero except at $k_z = \pm 0.5\,\pi/h_0$ (see details of calculation in Ref. [38]). The non-zero surface polarization $p_s$ ensures the existence of localized states on the "boundary of boundary", that is, the hinge, due to the surface-boundary correspondence [2,42]. Since the surface bandgaps completely overlap with the middle bulk band in a spectra range, hinge-localized states resulting from the nontrivial surface subspace can lie in the continuum of the other subspace [28], that is, the bulk bands, forming the BHICs but remaining intact.

*Acoustic realization and characterization of the HOWSM.*—To confirm our predictions aforementioned, we now implement the model of the HOWSM in a PC. As shown in Fig. 2(a), the 3D-printed sample is a parallelogram-shaped prism consisting of tube-connected cylindrical cavities. The cavities and the connecting tubes correspond respectively to the sites and hoppings in the model. Not to lose the generality, only the hinge close to the origin of the coordinates in the photos and its two adjacent surfaces are considered. Figure 2(b) gives the unit cell with $a_0 = 35.00$ mm and $h_0 = 11.82$ mm, the cylindrical cavities and connecting tubes are all filled with air. Note that the PC sample is terminated by hard wall but the integrity of the unit cells on the boundary is retained, as shown in the inset in Fig. 2(a). The boundary cavities, with half the length of the inner ones, play a vital role in emulating the open boundary condition in the tight-binding model [11,44]. Identical cavities, A (red), B (turquoise), and C (blue), with the dipolar modes chosen to map the zero-energy sites, are horizontally placed in a unit cell. The tubes [orange, silver, and black tubes in Fig. 2(b)] mimic three different hoppings. Since it demands simultaneously the positive and negative couplings for the desiring ideal HOWSM in the tight-binding model, straight (black tubes) and tilted (orange and silver tubes) connections are utilized according to the acoustic field distribution of dipolar model [13]. In Fig. 2(c), we present the simulated bulk band structure, a pair of ideal unconventional TWCs with oppositely charged WPs appear at 8.50 kHz and 10.06 kHz, agreeing well with the tight-binding model.

Then we numerically and experimentally demonstrate the topological surface states on the XZ surface of our PC. Figure 2(d) illustrates the experimental measurements. A broadband point sound source (white star) is fixed at the right side of the XZ surface and a microphone is inserted into the boundary cavities to probe the acoustic signals. The measured surface dispersions (colormaps) for various $k_z$ in Fig. 2(e), obtained from the Fourier transformation of the pressure field, agree well with the simulated surface states (yellow lines). The intensities of the measured surface dispersions around the frequencies of the quadratic WPs are weaker than the others due to the flatter local surface dispersions. Both the simulated and experimental results commendably reproduce the predictions from the tight-binding model.

*Observation of the BHICs*.—To validate the existence of two types of hinge states, the BHICs and the HIGs, the experimental measurement is performed as shown in Fig. 3(a). To excite the hinge modes, a point source (white star) is inserted into the boundary cavity at the middle of the hinge. By Fourier transforming the measured pressure field on the hinge (blue dashed line), the hinge dispersion can be obtained, as shown in Fig. 3(b). We can see that the experimental results (colormap) for the BHICs and HIGs agree well with the simulations (cyan and red lines). Due to the nontrivial bulk polarization, the HIGs, as expected, appear in the ranges $|k_z| > 0.5 \pi/h_0$ for the first bulk bandgap, and $|k_z| < 0.5 \pi/h_0$ for the second one, identifying our PC as a HOWSM. What is more, the BHICs, embedded in the bulk modes but remaining intact, is adequately addressed in the surface bandgaps owing to the nontrivial surface polarization.

The BHICs decoupling from the spectrally coexisting bulk modes, generally have high Q factors when resonating to excitation. This can be reflected by the largest spectral intensity in the colormap in Fig. 3(b). In Fig. 3(c), we exemplify the eigenstates of the BHICs and HIGs as labeled in Fig. 3(b). The existence of BHICs and HIGs can also be confirmed by the response spectra of the acoustic wave propagating along the hinge in Fig. 3(d), measured by exciting and probing the boundary cavities in the same column (along $z$ direction, as marked in the insets) around the hinge (see Ref. [38]). The dark grey shaded curves show several peaks of pressure, consistent with the simulated eigenfrequency ranges of the hinge states highlighted by the cyan and red regions. These curves have a stronger intensity in the cyan region, representing the BHICs, while a weaker intensity in the red regions, representing the HIGs.

*Discussion.*—To summarize, we proposed a theoretical route to construct a HOWSM hosting the unique BHICs. Remarkably, these BHICs, embedded in the continuum of bulk bands but remaining localized at the hinge without coupling with the radiating bulk modes, emerge in the surface bandgaps, as a result of the nontrivial surface polarization, regardless of the bulk polarization. Besides, the HOWSM contains two ideal TWCs, facilitating the observations of the usual topological surface states and HIGs. All results are verified by numerical simulations and airborne sound experiments. Our work extends the scope of topological states with the BHICs, opens a new avenue towards exploring higher-order topological states without requiring for spectral isolation, and broadens the material systems possessing BICs. Our work may promote the development of high-Q acoustic devices in applications.

**Acknowledgements**

We thank M. Xiao and H. Wu for helpful discussions. This work is supported by the National Key R&D Program of China (Grant Nos. 2022YFA1404900, 2018YFA0305800), the National Natural Science Foundation of China (Grant Nos. 11890701, 11974262, 12004286, and 12104347) and the China Postdoctoral Science Foundation (Grant Nos. BX20200259, 2020TQ0233, 2020M682461, and 2020M682462).

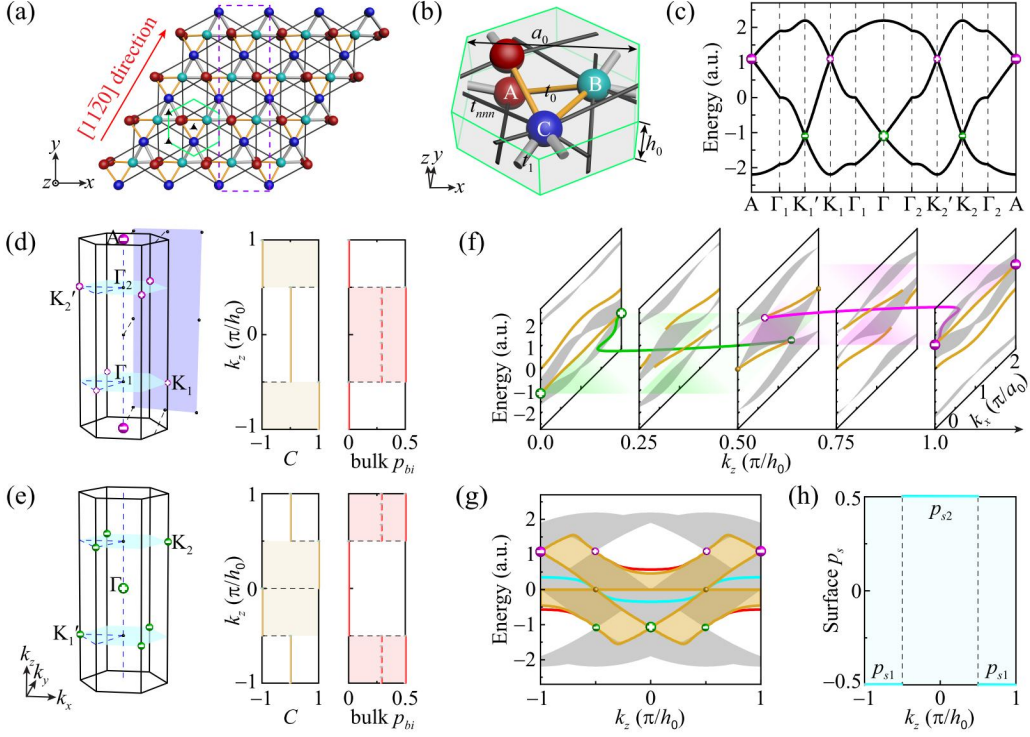

FIG. 1. Tight-binding model for the HOWSM with BHICs. (a) Top view of a 3D lattice of a parallelogram-shaped prism structure. Ribbon structure sketched by the purple dashed rectangle with periodic boundary conditions along the $x$ and $z$ directions is used to calculate the surface dispersions of the XZ surface. (b) Unit cell with three sites inside (colored spheres) coupled by nearest-neighbor hoppings $t_0$ (orange tubes) and $t_1$ (silver tubes), and next-nearest-neighbor hoppings $t_{nnn}$ (black tubes). (c) Bulk band structure along the blue dashed lines in the BZ in the left panel of (d) or (e). WPs in the two ideal TWCs are distinguished by green and magenta spheres respectively, with the big (small) spheres indicating the quadratic (charge-1) WPs, and "$\pm$" the sign of charge. (d),(e) Left panels: distributions of the WPs of the two TWCs in the BZ, respectively. Middle and right panels: $k_z$-dependent Chern number $C$ (middle) and bulk polarization $p_b$ (right, solid lines for $p_{bx}$ and dashed lines for $p_{by}$) in the corresponding bulk gaps. $\Gamma_{1(2)}$, $K_{1(2)}$ and $K'_{1(2)}$ are on the $k_z = \mp 0.5\,\pi/h_0$ plane. (f) Surface dispersions in terms of discretized $k_z$ on the XZ surface, the surface BZ is given in the back of the left panel of (d). Yellow lines denote the surface states, and the grey regions denote the bulk bands. Green and magenta spheres indicate the WPs, yellow spheres indicate the positions of the surface bandgap closing. Green and magenta lines represent the Fermi arc states at the two Weyl energies. (g) Hinge band structure along the $k_z$ direction. Cyan and red lines denote the BHICs and HIGs, respectively. Grey and yellow regions indicate the bulk and surface states. (h) Surface polarization $p_s$ in the $x$ direction obtained from the first surface band ($p_{s1}$) for $|k_z| > 0.5\,\pi/h_0$ and the second one ($p_{s2}$) for $|k_z| < 0.5\,\pi/h_0$. Throughout (c)-(h), $t_0 = 0.366$, $t_1 = 1$, and $t_{nnn} = -0.134$ are taken.

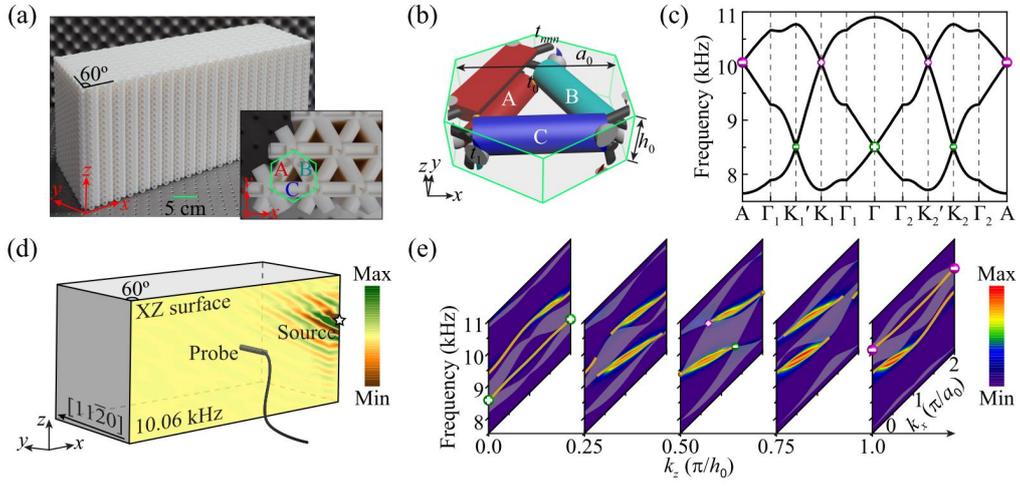

FIG. 2. Acoustic realization of the HOWSM and observation of the topological surface states. (a) Photo of the parallelogram-shaped prism PC sample. The inset gives the top view, with the green hexagon sketching a unit cell. (b) Unit cell of the PC, the colored regions are filled with air and bounded with hard boundary. (c) Simulated bulk band structure of the PC. (d) Experiment for surface states observation. The colormap represents the measured surface field at 10.06 kHz. (e) Measured surface dispersions (colormap) in terms of discretized $k_z$. Simulated bulk and surface states are also plotted, as denoted by the grey shadows and the yellow lines, respectively.

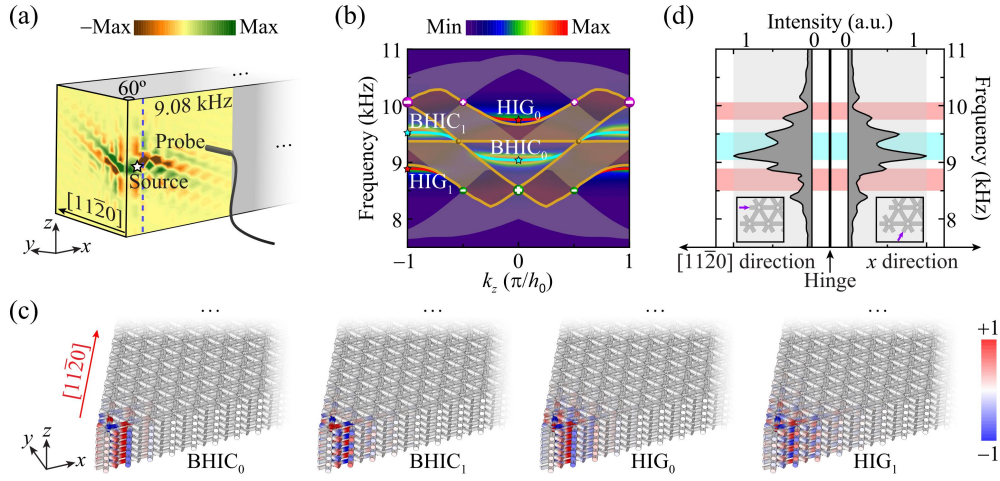

FIG. 3. Observation of the BHICs. (a) Experiment for hinge states observation. The colormaps represent the measured acoustic fields around the hinge at 9.08 kHz. (b) Measured hinge dispersion. The colormap denotes the experimental result. Cyan and red lines highlight the simulated BHICs and HIGs, respectively, compared with the simulated bulk (grey shadows) and surface (yellow regions) states. (c) Simulated eigenfields for hinge states, $BHIC_{0(1)}$ and $HIG_{0(1)}$, of $k_z = 0$ ($-\pi/h_0$), indicated by the colored stars in (b). (d) Measured response spectra at a site close to the hinge, seven sites above the source fixed in the same column, as indicated in the insets by the arrows. The cyan (red) regions label the eigenfrequency ranges of the BHICs (HIGs). The insets mark the exciting or probing columns by colored arrows.